# Evaluation of the Structural disorder of the protein FMR1 with Carbon Composition

Baby Jerald A. #1
Research Associate – RIIC,
Biocomputing Research Group,
Dayananda Sagar Institutions
Bangalore, India
[1]babyjerald@gmail.com

Dr. T.R. Gopalakrishnan Nair*2
ARAMCO Endowed Chair –Technology,
PMU, KSA.
Vice President – RIIC,
Dayananda Sagar Institutions
Bangalore, India
[2]trgnair@gmail.com

Dr. Ekambaram Rajasekaran*3
Department of Bioinformatics,
School of biotechnology
Karunya university
Coimbatore, India
[3]ersekaran@gmail.com

*Abstract*: Ever since the disorder of proteins is the main cause for many diseases. As compared with other disorders, the major reason that causes disease is of structural inability of many proteins. The potentially imminent availability of recent datasets helps one to discover the protein disorders, however in majority of cases, the stability of proteins depend on the carbon content. Addressing this distinct feature, it is possible to hit upon the carbon distribution along the sequence and can easily recognize the stable nature of protein. There are certain reported mental disorders which fall in to this category. Regardless, such kind of disorder prone protein FMR1p (Fragile X mental retardation 1 protein) is identified as the main cause for the disease Fragile X syndrome. This paper deals with the identification of defects in the FMR1 protein sequence considering the carbon contents along the sequence. This attempt is to evaluate the stability of proteins, accordingly the protein disorders in order to improvise the certain Biological functions of proteins to prevent disease. The transition of the disorder to order protein involves careful considerations and can be achieved by detecting the unstable region that lacks hydrophobicity. This work focuses the low carbon content in the FMR1 protein so as to attain the stable status in future to reduce the morbidity rate caused by Fragile X syndrome for the society.

## I. INTRODUCTION:

Fragile X syndrome evolves its greatest threat to human kind with severe mental retardation and in general known as genomic syndrome due to the expansion of trinucleotide gene sequence (CGG) within X chromosome [1, 2, 3, 4] What if the gene expands? And what could be the effects? To address these issues, it is been identified that the CGG repeat get amplified just before the coding region and intensely disrupts the final end product FMR1 protein, thus the expression of FMR1 protein is hindered [5, 6, 7]. This results in macroorchidism (enlargement of the testicles), large ears, prominent jaw, and high-pitched, jocular speech [8, 9, 10, 11, 12, 13, 14, and 15]. People with fragile X syndrome is believed to have 55 to 200 CGG repeats whereas the normal individuals have 6-54 CGG sequences [16]. Ultimately it results in methylation of FMR gene and the biological function of the FMR1 protein like translational efficiency and/or trafficking of certain mRNAs is disturbed and remains unstable [17, 18]. It is intended that the unstable protein lost its nature to fold itself to involve in biological process. Current research shows that carbon content along the sequence plays important role in maintaining the stability of the proteins and it has been proved 31.45% carbon content prefers to promote the stable nature of protein [19]. With this analysis, the disordered region of the protein sequences can be identified for further research programs. Henceforth the development of robust scientifically informed guidance on how best to improve the stable nature of protein need to be implemented for better function of FMR1 protein to depose the Down syndrome. In this work we attempt to show the nonstable region of the FMR1 protein based on carbon composition that is responsible for the misfolding of proteins.

## II. RESEARCH BACKGROUND:

Many research works are based on the folded confirmation of the proteins so as to retrieve the tertiary structure and to carry out specific biological function. Nonetheless, most of the proteins remain unfolded and thus reported as disordered proteins. As a result, it involves complex scientific, ethical and political considerations such as whether and how best to deal the protein disorder [20]. The role of FMR1 protein results in many functions as stated above. Kwon et al., 2001 in his work says the expression of FMR1 protein inherently increases the activation of parietal lobe and also involved in early brain development. It also accelerates the regulation of protein synthesis in synapses [17]. The estimation of FMR1 protein expression within neurons is been done by Yucui Chen et al., 2010. All these works adds weight to the importance of FMR1 protein.

All previous studies have shown that FMR1 gene alone is considered unlikely to pose unique risks and also FMR1p found to accelerate all normal neural functions [22, 23]. Such gene expansion will disrupts the expression of FMR1p and ends up in synaptic abnormalities. As a result many functions of synaptic proteins are up or down regulated and ultimately affect the neurotransmission and ends up in some neuronal disorders [24]. Such defect in FMR1p expression is found likely to be reporting some adverse events like dendritic abnormalities [25]. Finally the main cause of Fragile X syndrome is found to be associated with the defects in FMR1 protein expression.

## III. Methodology:

The goal is to understand the molecular level of FMR1 protein. In general the folding properties of proteins depend on non covalent interactions. The rich and scarce distribution of carbon along the sequence contributes the folding nature of proteins. Accordingly, it confers the stability of proteins [26]. It is clear that considerable inherent variability in the carbon content affects the structure of proteins. For this reason, it is important to calculate the carbon level to encounter the disorder protein. The ideology is that to predict the carbon content with applicable C program called CARBANA [19]

The carbon prediction may provide valuable insights in Medical research and the initial dataset for this analysis can be accessed through online available databases. Here the disorder protein of FMR1 is accessed with DisProt (www.disprot.org). Similar databases are Disopred2 [27], OnD-CRF [28], FoldIndex [29], GeneSilico Metadisoreder [30], MFDp [31]. The disorder FMR1 protein was retrieved and its disport ID is DP00134 and the Uniprot ID is Q06787.

Fasta Sequence of FMR1 protein:

>DisProt|DP00134|uniprot|Q06787|unigene|Hs.103183|sp|FMR1_HUMAN|gi|544328 #281-422:Protein-mRNA binding #516-632:Protein-mRNA binding
MEELVVEVRGSNGAFYKAFVKDVHEDSITVAFENNWQPDRQIPFHDVRFP
PPVGYNKDINESDEVEVYSRANEKEPCCWWLAKVRMIKGEFYVIEYAACDA
TYNEIVTIERLRSVNPNKPATKDTFHKIKLDVPEDLRQMCAKEAAHKDFKK
AVGAFSVTYDPENYQLVILSINEVTSKRAHMLIDMHFRSLRTKLSLIMRNEE
ASKQLESSRQLASRFHEQFIVREDLMGLAIGTHGANIQQARKVPGVTAIDLD
EDTCTFHIYGEDQDAVKKARSFLEFAEDVIQVPRNLVGKVIGKNGKLIQEIV
DKSGVVRVRIEAENEKNVPQEEEIMPPNSLPSNNSRVGPNAPEEKKHLDIKE
NSTHFSQPNSTKVQRVLVASSVVAGESQKPELKAWQGMVPFVFVGTKDSIA
NATVLLDYHLNYLKEVDQLRLERLQIDEQLRQIGASSRPPPNRTDKEKSYVT
DDGQGMGRGSRPYRNRGHGRRGPGYTSGTNSEASNASETESDHRDELSDW
SLAPTEEERESFLRRGDGRRRGGGGRGQGGRGRGGGFKGNDDHSRTDNRP
RNPREAKGRTTDGSLQIRVDCNNERSVHTKTLQNTSSEGSRLRTGKDRNQK
KEKPDSVDGQQPLVNGVP

CARBANA (www.rajasekaran.net.in/tools/carbana.html) has window size limit and the chosen window size is 700 as its sequence length is 632. The DisProt shows 41% disorder in FMR1 Protein. It is analyzed with its carbon composition and has been visualized (Fig 1).

## IV. Results and Discussion:

FMR1 gene and FMR1 protein intricately get involved in Fragile X syndrome. The initiative of this work focused on disorder portion of the FMR protein sequence. It gives the analysis to certain extent with which the unstable portion of the protein is predicted with the help of C program CARBANA. It tried to cover the disorder portion to the maximum by estimating the carbon distribution along the sequence. The misfolding and unfolding region of protein depends on carbon content and protein acquires its stable form with 31.45% carbon (Sneha et al., 2011). If any change in the carbon content happens, it leads to the disorder. The prediction of disorder portion helps to figure out the root cause for fragile X syndrome.

Figure 1 represents the disorder region of the protein and at the molecular level; we tried to show the carbon distribution with the provided input. As a result the amino acid position and its carbon composition are predicted and visualized (Table 1).

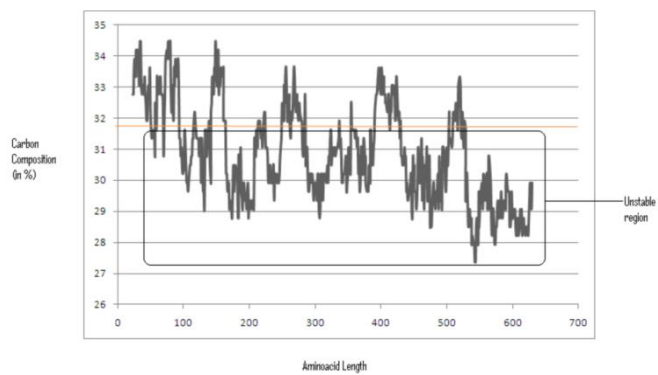

Figure 1: Plot of carbon content in FMR1 protein showing the disorder region. The portion below 31.45 % lacks stability.

## V. Conclusion:

Overall analysis provides an evidence of disorder portion of the protein FMR1. This striking observation of this analysis helps to identify the disorder in FMR1 protein. Our finding of protein sequence below the normal carbon content (31.45%) indicates the protein can misregulate and that approximates the reported gene disruption. Our work greatly expands to detect the possible cause of malfunctioning of the protein in causing fragile X syndrome. It provides potential insights in to underlying mechanism such as the failure of FMR1 protein expression due to disorder nature and unfits to fold itself to attain certain conformations to perform its biological function. Further studies using recent technologies can explore the translation of disorder to order protein.

VII. APPENDICES

Table 1: Carbon Distribution in FMR1 protein

| Aminoacid | Carbon composition (in %) | Aminoacid | Carbon composition (in %) | Aminoacid | Carbon composition (in %) |
|---|---|---|---|---|---|
| 23 | 32.76 | 221 | 31.62 | 429 | 31.34 |
| 24 | 32.76 | 222 | 31.91 | 430 | 31.91 |
| 25 | 33.62 | 223 | 32.19 | 431 | 30.77 |
| 26 | 33.9 | 224 | 32.19 | 432 | 30.77 |
| 28 | 33.33 | 225 | 31.05 | 433 | 30.48 |
| 28 | 34.19 | 226 | 31.05 | 434 | 30.48 |
| 30 | 34.19 | 227 | 31.05 | 435 | 30.48 |
| 31 | 34.19 | 228 | 30.77 | 436 | 31.05 |
| 32 | 33.62 | 229 | 29.91 | 437 | 30.77 |
| 33 | 33.05 | 230 | 29.91 | 438 | 30.77 |
| 34 | 34.47 | 231 | 30.48 | 438 | 30.77 |
| 35 | 34.47 | 232 | 29.91 | 439 | 29.63 |
| 36 | 33.05 | 232 | 30.48 | 440 | 29.91 |
| 37 | 32.76 | 233 | 29.91 | 441 | 29.34 |
| 38 | 33.05 | 235 | 29.91 | 443 | 29.63 |
| 39 | 33.05 | 236 | 30.2 | 444 | 29.63 |
| 40 | 32.76 | 237 | 30.48 | 444 | 29.63 |
| 41 | 33.33 | 238 | 29.34 | 445 | 29.91 |
| 42 | 32.76 | 239 | 30.2 | 446 | 30.2 |
| 43 | 32.48 | 240 | 29.91 | 447 | 29.06 |
| 44 | 31.91 | 242 | 30.2 | 448 | 28.77 |
| 45 | 31.91 | 243 | 29.91 | 449 | 30.77 |
| 46 | 32.76 | 245 | 29.91 | 450 | 29.91 |
| 47 | 33.05 | 246 | 30.2 | 451 | 30.48 |
| 48 | 32.76 | 247 | 30.77 | 451 | 29.63 |
| 49 | 33.62 | 249 | 31.34 | 453 | 29.63 |
| 50 | 32.19 | 249 | 31.34 | 454 | 29.06 |
| 51 | 31.62 | 250 | 32.48 | 455 | 29.34 |
| 52 | 31.34 | 251 | 32.48 | 456 | 30.2 |
| 53 | 31.62 | 253 | 31.91 | 456 | 30.48 |
| 55 | 31.34 | 253 | 33.05 | 457 | 31.05 |
| 56 | 31.34 | 254 | 32.48 | 458 | 29.91 |
| 57 | 30.77 | 255 | 33.33 | 460 | 29.91 |
| 57 | 32.48 | 256 | 33.62 | 462 | 31.34 |
| 59 | 31.62 | 258 | 31.91 | 463 | 29.34 |
| 60 | 33.33 | 259 | 32.19 | 463 | 30.77 |
| 61 | 33.05 | 261 | 32.76 | 465 | 29.91 |
| 62 | 32.76 | 261 | 32.19 | 466 | 29.06 |
| 63 | 33.05 | 263 | 31.34 | 467 | 30.48 |
| 65 | 33.33 | 264 | 32.19 | 468 | 30.48 |
| 66 | 32.76 | 265 | 31.91 | 469 | 30.48 |
| 67 | 32.76 | 266 | 32.19 | 470 | 31.05 |
| 68 | 32.76 | 268 | 33.62 | 471 | 29.34 |
| 69 | 32.48 | 269 | 33.33 | 472 | 29.63 |
| 70 | 30.77 | 270 | 32.48 | 473 | 29.63 |
| 71 | 31.91 | 271 | 32.19 | 474 | 30.77 |
| 72 | 32.76 | 272 | 32.48 | 475 | 28.49 |
| 73 | 33.9 | 273 | 32.76 | 476 | 28.77 |
| 75 | 34.19 | 273 | 32.48 | 477 | 28.49 |
| 76 | 33.9 | 275 | 31.91 | 478 | 29.06 |
| 76 | 33.9 | 276 | 32.48 | 480 | 29.91 |
| 77 | 34.47 | 277 | 32.19 | 481 | 29.63 |
| 79 | 33.9 | 279 | 31.91 | 483 | 29.06 |
| 80 | 34.47 | 280 | 31.91 | 484 | 29.63 |
| 81 | 33.05 | 281 | 31.05 | 485 | 29.34 |
| 82 | 32.19 | 282 | 30.77 | 487 | 30.2 |
| 83 | 32.48 | 283 | 31.62 | 488 | 29.06 |
| 83 | 33.05 | 284 | 31.34 | 489 | 30.77 |
| 85 | 31.91 | 285 | 32.76 | 490 | 29.34 |
| 86 | 33.62 | 286 | 30.48 | 491 | 29.91 |
| 87 | 33.05 | 287 | 30.77 | 491 | 29.63 |
| 87 | 33.33 | 288 | 31.05 | 492 | 29.91 |
| 88 | 33.9 | 289 | 30.2 | 493 | 29.06 |
| 89 | 33.33 | 290 | 29.63 | 494 | 31.05 |
| 90 | 33.62 | 291 | 29.63 | 497 | 30.77 |
| 91 | 33.62 | 292 | 30.2 | 498 | 29.91 |
| 92 | 33.9 | 294 | 30.2 | 499 | 29.63 |
| 93 | 33.05 | 294 | 29.91 | 500 | 30.48 |
| 94 | 31.34 | 295 | 30.2 | 500 | 31.05 |
| 95 | 31.34 | 296 | 30.2 | 502 | 30.48 |
| 96 | 30.77 | 298 | 29.34 | 504 | 32.19 |
| 97 | 31.05 | 298 | 29.91 | 504 | 31.34 |
| 98 | 30.48 | 299 | 29.63 | 506 | 31.34 |
| 99 | 30.2 | 300 | 29.91 | 508 | 31.05 |
| 101 | 30.48 | 301 | 29.91 | 509 | 31.34 |
| 102 | 31.62 | 303 | 29.34 | 510 | 32.19 |
| 104 | 30.48 | 304 | 30.2 | 511 | 32.19 |
| 105 | 29.91 | 305 | 30.2 | 513 | 31.91 |
| 106 | 29.91 | 306 | 29.06 | 514 | 32.48 |
| 107 | 29.63 | 307 | 28.77 | 516 | 32.19 |
| 108 | 29.91 | 308 | 29.06 | 517 | 31.62 |
| 109 | 30.2 | 309 | 30.2 | 518 | 33.05 |
| 110 | 30.48 | 310 | 30.2 | 520 | 33.33 |
| 111 | 30.48 | 312 | 29.34 | 521 | 33.05 |
| 112 | 30.77 | 313 | 30.2 | 522 | 31.62 |
| 113 | 30.77 | 314 | 30.2 | 523 | 31.34 |
| 114 | 31.34 | 314 | 30.2 | 524 | 32.19 |
| 115 | 31.34 | 315 | 30.2 | 525 | 31.62 |
| 115 | 31.62 | 317 | 29.91 | 526 | 31.05 |
| 117 | 32.19 | 317 | 29.91 | 527 | 31.91 |
| 118 | 31.91 | 319 | 30.48 | 529 | 31.62 |
| 119 | 31.62 | 320 | 29.91 | 529 | 29.34 |
| 120 | 31.34 | 321 | 30.77 | 531 | 29.63 |
| 121 | 31.34 | 322 | 30.77 | 531 | 30.2 |
| 122 | 31.34 | 324 | 31.62 | 533 | 28.49 |
| 123 | 31.34 | 325 | 31.05 | 534 | 28.49 |
| 124 | 30.77 | 325 | 30.48 | 535 | 29.06 |
| 126 | 31.34 | 326 | 30.77 | 536 | 28.77 |
| 127 | 29.91 | 327 | 30.77 | 538 | 28.77 |
| 128 | 31.34 | 328 | 31.05 | 538 | 27.92 |
| 129 | 30.48 | 329 | 31.05 | 539 | 28.21 |
| 130 | 29.63 | 330 | 31.05 | 541 | 27.92 |
| 131 | 29.63 | 331 | 30.77 | 541 | 27.92 |
| 132 | 29.06 | 333 | 31.05 | 542 | 27.92 |
| 133 | 31.34 | 334 | 31.05 | 543 | 27.35 |
| 133 | 31.62 | 334 | 31.05 | 544 | 27.35 |
| 134 | 31.62 | 336 | 31.34 | 545 | 28.49 |
| 135 | 30.77 | 337 | 31.91 | 546 | 28.77 |
| 136 | 31.05 | 338 | 31.05 | 548 | 27.92 |
| 137 | 31.05 | 339 | 31.34 | 549 | 29.63 |
| 139 | 31.91 | 340 | 30.77 | 549 | 28.49 |
| 140 | 30.77 | 341 | 29.91 | 550 | 28.77 |
| 141 | 30.2 | 342 | 29.91 | 551 | 29.34 |
| 142 | 29.91 | 343 | 30.2 | 554 | 29.91 |
| 142 | 32.19 | 344 | 30.2 | 555 | 29.06 |
| 143 | 32.48 | 345 | 30.2 | 556 | 30.2 |
| 144 | 33.33 | 347 | 30.77 | 558 | 29.34 |
| 146 | 32.76 | 348 | 29.63 | 559 | 29.91 |
| 147 | 33.62 | 349 | 29.91 | 562 | 30.2 |
| 148 | 33.62 | 350 | 30.2 | 562 | 29.06 |
| 149 | 34.47 | 351 | 31.05 | 564 | 29.06 |
| 151 | 33.05 | 353 | 30.48 | 564 | 30.77 |
| 152 | 33.33 | 354 | 30.2 | 567 | 29.91 |
| 153 | 34.19 | 355 | 32.48 | 568 | 28.77 |
| 154 | 32.76 | 356 | 31.62 | 569 | 28.21 |
| 155 | 33.33 | 357 | 31.62 | 570 | 28.77 |
| 155 | 33.33 | 359 | 31.62 | 571 | 29.06 |
| 156 | 33.05 | 360 | 31.62 | 573 | 27.92 |
| 157 | 33.62 | 361 | 31.34 | 574 | 27.92 |
| 159 | 33.33 | 362 | 31.62 | 575 | 28.49 |
| 161 | 33.62 | 363 | 31.34 | 576 | 28.49 |
| 161 | 31.91 | 364 | 31.05 | 577 | 29.06 |
| 163 | 31.91 | 365 | 31.34 | 578 | 29.34 |
| 164 | 31.91 | 366 | 30.2 | 580 | 29.06 |
| 165 | 30.2 | 367 | 30.77 | 580 | 28.77 |
| 166 | 30.2 | 368 | 29.63 | 581 | 29.34 |
| 167 | 29.63 | 369 | 29.91 | 582 | 29.06 |
| 168 | 29.63 | 370 | 31.05 | 583 | 29.63 |
| 169 | 29.91 | 371 | 30.77 | 584 | 28.77 |
| 170 | 29.06 | 372 | 30.48 | 585 | 28.77 |
| 171 | 29.34 | 373 | 30.2 | 586 | 29.06 |
| 172 | 29.06 | 374 | 31.34 | 587 | 29.34 |
| 173 | 29.06 | 376 | 30.77 | 588 | 29.34 |
| 174 | 28.77 | 377 | 30.77 | 590 | 29.06 |
| 175 | 29.63 | 377 | 29.91 | 590 | 29.91 |
| 176 | 30.48 | 379 | 29.63 | 591 | 30.2 |
| 177 | 29.63 | 380 | 29.63 | 593 | 29.63 |
| 178 | 30.2 | 381 | 29.63 | 595 | 29.63 |
| 179 | 30.48 | 382 | 30.48 | 596 | 28.49 |
| 181 | 30.2 | 384 | 29.91 | 597 | 28.49 |
| 182 | 29.91 | 384 | 31.62 | 598 | 28.77 |
| 183 | 28.77 | 385 | 31.62 | 599 | 28.77 |
| 184 | 28.77 | 386 | 30.77 | 599 | 29.63 |
| 184 | 30.77 | 387 | 30.48 | 600 | 28.77 |
| 185 | 30.2 | 388 | 30.48 | 602 | 29.06 |
| 186 | 29.63 | 389 | 31.05 | 603 | 29.06 |
| 187 | 31.05 | 390 | 31.34 | 604 | 29.06 |
| 188 | 29.63 | 391 | 32.19 | 605 | 28.49 |
| 189 | 29.34 | 393 | 32.76 | 606 | 28.21 |
| 190 | 29.06 | 394 | 32.76 | 607 | 28.77 |
| 191 | 29.91 | 395 | 33.05 | 608 | 28.77 |
| 192 | 29.91 | 397 | 33.62 | 609 | 28.21 |
| 193 | 29.63 | 398 | 33.05 | 611 | 28.49 |
| 194 | 29.34 | 400 | 32.76 | 611 | 28.49 |
| 195 | 29.34 | 401 | 32.76 | 612 | 29.06 |
| 196 | 29.06 | 401 | 33.62 | 613 | 28.49 |
| 196 | 29.06 | 403 | 33.05 | 614 | 29.06 |
| 198 | 29.91 | 404 | 32.76 | 615 | 28.21 |
| 199 | 29.06 | 405 | 33.33 | 617 | 28.77 |
| 199 | 28.77 | 405 | 32.76 | 618 | 28.49 |
| 200 | 29.34 | 407 | 32.76 | 620 | 28.21 |
| 201 | 29.34 | 407 | 32.48 | 621 | 28.49 |
| 202 | 29.34 | 408 | 32.48 | 623 | 28.49 |
| 203 | 29.06 | 410 | 32.19 | 624 | 28.21 |
| 204 | 29.34 | 411 | 32.48 | 625 | 28.77 |
| 205 | 29.34 | 412 | 32.76 | 626 | 29.34 |
| 206 | 29.34 | 413 | 31.91 | 626 | 29.34 |
| 207 | 29.06 | 414 | 31.62 | 627 | 29.91 |
| 208 | 31.34 | 415 | 32.19 | 629 | 29.06 |
| 209 | 31.05 | 416 | 32.19 | 630 | 29.91 |
| 210 | 30.77 | 417 | 33.05 | | |
| 211 | 31.62 | 419 | 32.76 | | |
| 213 | 31.05 | 419 | 33.05 | | |
| 213 | 31.62 | 421 | 32.76 | | |
| 214 | 31.91 | 422 | 31.91 | | |
| 215 | 31.62 | 423 | 32.76 | | |
| 216 | 31.91 | 424 | 33.33 | | |
| 218 | 31.34 | 426 | 32.19 | | |
| 219 | 31.62 | 427 | 31.91 | | |
| 220 | 31.05 | 428 | 32.19 | | |